\documentclass[10pt]{article}
\usepackage[utf8]{inputenc}
\usepackage[T1]{fontenc}
\usepackage[top=1in,bottom=1in,left=1in,right=1in]{geometry}
\usepackage{authblk}
\usepackage{amsmath,amsfonts,amssymb,amsthm}

\usepackage{tikz-cd}

\numberwithin{equation}{section}

\newcommand{\R}{\mathbb{R}}
\newcommand{\C}{\mathbb{C}}

\newcommand{\bin}[2]{\left(\begin{matrix} #1 \\ #2 \end{matrix}\right)}
\newcommand{\up}{\Delta}
\newcommand{\dn}{\nabla}
\newcommand{\pol}{\mathcal{P}}
\newcommand{\y}{\lambda(x)}
\newcommand{\uy}{2x+\gamma+\delta+2}
\newcommand{\dy}{2x+\gamma+\delta}
\newcommand{\al}{\theta}

\title{The Heun-Racah and Heun-Bannai-Ito algebras}
 \author[1]{Geoffroy Bergeron}
 \author[2]{Nicolas Cramp\'e}
 \author[4]{Satoshi Tsujimoto}
 \author[1]{Luc Vinet}
 \author[3]{Alexei Zhedanov}
 \affil[1]{Centre de recherches math\'ematiques, Universit\'e de Montr\'eal, P.O. Box 6128, Centre-ville Station, Montr\'eal, Canada H3C 3J7}
 \affil[2]{Institut Denis-Poisson CNRS/UMR 7013, Universit\'e de Tours - Universit\'e d'Orl\'eans, Parc de Grandmont, 37200 Tours, France}
 \affil[3]{School of Mathematics, Renmin University of China, Beijing 100872, China}
 \affil[4]{Department of Applied Mathematics and Physics, Graduate School of Informatics, Kyoto University, Kyoto, 606-8501, Japan}

\date{\today}
\begin{document}
\maketitle
\thispagestyle{empty}
\hrule
\begin{abstract}\noindent
This paper introduces and studies the Heun-Racah and Heun-Bannai-Ito algebras abstractly and establishes the relation between these new algebraic structures and generalized Heun-type operators derived from the notion of algebraic Heun operators in the case of the Racah and Bannai-Ito algebras.
\end{abstract} 
\hrule

\section{Introduction}
Systematic generalizations of the standard Heun operator were recently introduced in \cite{Grunbaum2018}. In this approach, an algebraic Heun operator is associated to any bispectral pair \cite{Grunbaum2001} of operator. To do so, one considers the algebra generated by such a pair and defines the algebraic Heun operator as the generic bilinear combination of the operators in the bispectral pair. Such a construction has proved useful \cite{Crampe2019} in the theory of time and band limiting \cite{Slepian1983,Landau1985} as well as in the study of entanglement in fermionic chains \cite{Crampe2019a,Crampe2020} and the related algebraic structures are now being studied. Of particular interest are the algebraic Heun operators constructed from the bispectral problems arising in the theory of orthogonal polynomials. This paper aims to introduce the Heun-Racah and the Heun-Bannai-Ito algebras and the associated algebraic Heun operators. Let us first review the recent results in this setting.

Orthogonal polynomials of the Askey scheme \cite{Koekoek2010} are naturally associated to bispectral pairs. Indeed, all these polynomials are the eigenfunctions of both a three-term recurrence operator $X$, acting on the degree, and a differential or difference operator $Y$, acting on the variable. Thus, $X$ and $Y$ form a bispectral pair. One then considers the algebra generated by the pair $X$ and $Y$ in either the variable or the degree representation. The resulting structures are quadratic algebras canonically associated to each polynomials and after which they are named. These provide an algebraic approach to the theory of orthogonal polynomials as one recovers the properties of the corresponding polynomial through the representation theory of the associated algebra. Furthermore, the structure of the Askey scheme with other families of polynomials appearing as limits and specializations of higher polynomials is reflected algebraically through specializations and contractions. It is in this algebraic setting that one defines the associated algebraic Heun operator as
\begin{align}\label{wdefalg}
 W = \tau_1 XY + \tau_2 YX + \tau_3 X + \tau_4 Y + \tau_0,
\end{align}
where $\tau_{i} \in \C$ for $i=0,1,2,3,4$ are arbitrary parameters. Following the naming convention of the algebras associated to polynomials in the Askey scheme, operators constructed as in \eqref{wdefalg} will be referred to in terms of the name of the associated polynomials. For instance, if one considers the Racah polynomials and the associated Racah algebra, the algebraic Heun operator $W$ in this setting is referred to as the Heun-Racah operator.

The algebraic Heun operator can be seen to generalize the Heun operator in many ways. It corresponds \cite{Grunbaum2017} to the standard Heun operator when constructed from the Jacobi algebra. In this case, $Y$ is the hypergeometric operator while $X$ is multiplication by the variable. Moreover, the standard Heun operator can be characterized as the most general second order differential operator that sends polynomials to polynomials one degree higher. The algebraic Heun operators constructed from the other entries in the Askey scheme are observed to satisfy analogs of this property. The Heun-Hahn operator introduced in \cite{Vinet2019} corresponds to the most general second-order difference operator on the uniform grid that sends polynomials on the grid to polynomials one degree higher. Likewise, the most general degree raising second order $q$-difference operator has been identified as an algebraic Heun operator in \cite{Baseilhac2019} and seen to correspond to a $q$-deformed Heun operator as considered in \cite{Takemura2018}. Finally, a $q$-Heun operator on the Askey grid was introduced in \cite{Baseilhac2018}. Such operators has been diagonalized recently in \cite{Baseilhac2019a} by a generalization of the algebraic Bethe ansatz, called the modified algebraic Bethe ansatz, introduced to solve spin chains with generic boundaries in \cite{Belliard2013,Belliard2015}.

A relevant property of the algebraic Heun operator $W$ is characterized as follows. Taking the realization where $X$ acts as the recurrence operator and $Y$, as multiplication by the eigenvalue, it is easily seen that $W$, as given by \eqref{wdefalg}, will be tridiagonal on the corresponding orthogonal polynomials. Correspondingly, in the finite-dimensional setting where the bispectral pair is taken to be a Leonard pair, it was proven in \cite{Nomura2007} that all tridiagonal operators take the form of \eqref{wdefalg}. There is a manifest relation between the construction \eqref{wdefalg} of $W$ and the tridiagonalization approach \cite{Ismail2012,Genest2016} to the study of orthogonal polynomials. From an algebra standpoint, tridiagonalization amounts to the construction of morphisms between the algebras associated to the polynomials of the Askey scheme. By considering the subalgebra generated by certain $W$ and either $X$ or $Y$, an embedding of the algebra of higher polynomials into the algebra of lower ones can be found. Naturally, this identification cannot be made when $W$ is constructed from the algebra of a polynomial sitting at the top of the Askey scheme as the resulting structures lie beyond the algebraic framework of the Askey scheme. In general, one is lead to the study of algebras, referred to in terms of the underlying polynomials, for instance, if $X$ and $Y$ are the generators of the Racah algebra, the algebra generated by the pair $X,W$ or $Y,W$ shall be called the Heun-Racah algebra. Characterizations of these Heun algebras have been done in \cite{Baseilhac2018} for the Heun-Askey-Wilson algebra, in \cite{Vinet2019} for the Heun-Hahn algebra and in \cite{Crampe2019} for the Heun algebras of the Lie type which encompasses the cases of the Krawtchouk, Meixner, Meixner-Pollaczek, Laguerre and Charlier polynomials. In this paper, a similar characterization is made of the Heun-Racah and the Heun-Bannai-Ito algebras.

The presentation is as follows. In section \ref{racah}, the Racah algebra is introduced and some key results are reviewed. The Heun-Racah operator is then constructed in section \ref{heun-racah-operator} as the most general second degree difference operator on the Racah grid that sends polynomials to polynomials one degree higher. As the algebraic Heun operator \eqref{wdefalg} of the Racah type is a bilinear combination of the Racah operator, it can be realized as a difference operator on the Racah grid using the canonical realization of the Racah algebra. This realization is shown to coincide with the Heun-Racah operator. Section \ref{heun-racah-algebra} defines the Heun-Racah algebra abstractly and gives the conditions on the parameters that define a specialization that has an embedding in the Racah algebra. This embedding allows a realization of this specialized Heun-Racah algebra to be induced from the canonical realization of the Racah algebra. The realization thus obtained is seen to be in terms of the Heun-Racah operator. Similarly, after reviewing the Bannai-Ito algebra in section \ref{bannai-ito}, the Heun-Bannai-Ito algebra, together with the Heun-Bannai-Ito operator, are introduced in section \ref{heun-bannai-ito}. It is shown that there is a specialization, obtained by imposing conditions on the parameters, that can be embedded in the Bannai-Ito algebra and realized in terms of the Heun-Bannai-Ito operator. A brief conclusion follows.

\section{The Racah algebra}\label{racah}
The Racah algebra $\mathcal{R}$ is the quadratic algebra defined as a unital associative algebra over $\C$ that is generated by $\{K_{1},K_{2},K_{3}\}$ with the following relations
\begin{align}
 [K_1,K_2] &= K_3,\nonumber \\
 [K_2,K_3] &= a_1 \{K_1,K_2\} + a_2 K_2^2 + bK_2 + c_1 K_1 + d_1,\label{racah-rel}\\
 [K_3,K_1] &= a_1 K_1^2 + a_2\{K_1,K_2\} + bK_1 + c_2 K_2 + d_2,\nonumber
\end{align}
with $a_{i}$, $c_{i}$ and for $i=1,2$ being arbitrary parameters in $\R$ and where $b$ and $d_{i}$ for $i=1,2$ are central elements. Throughout this paper, $[A,B]\equiv AB-BA$, $\{A,B\}\equiv AB+BA$ and $I$ denote, respectively, the commutator, the anti-commutator and the identity element. To simplify notations and allow the construction of a Poincaré-Birkhoff-Witt type basis, a third generator $K_{3}$ is introduced in this presentation, although it is not algebraically independent from the others. A relevant observation is that the relations \eqref{racah-rel} are fixed \cite{Genest2014} by considering the most general quadratic associative algebra generated by an independent pair of generators that admits ladder representations \cite{Genest2014a} and demanding compatibility with the following Jacobi identity
\begin{align}\label{jacobi-racah}
[K_{1},[K_{2},K_{3}]] + [K_{3},[K_{1},K_{2}]] + [K_{2},[K_{3},K_{1}]] = 0.
\end{align}
The Racah algebra is known to admit a cubic Casimir element $C$. In the above presentation, this central element is given by
\begin{multline}\label{racah-casimir}
C = a_{1} \{ K_{1}^{2},K_{2}\} + a_{2} \{ K_{1}, K_{2}^{2} \} + (a_{1} a_{2} + b) \{ K_{1},K_{2} \}\\
+ (a_{1}^{2}+c_{1}) K_{1}^{2} + (a_{2}^{2} + c_{2}) K_{2}^{2} + K_{3}^{2} + (a_{1} b + 2 d_{1}) K_{1} + (a_{2} b + 2 d_{2}) K_{2}.
\end{multline}

\subsection{The equitable presentation of the Racah algebra}
The Racah algebra is known to admit another presentation that displays explicitly the permutation symmetry of the generators. The relation between this second presentation and the one given in \eqref{racah-rel} is built \cite{Genest2014a} upon the reduced form of the Racah algebra that admits only three free parameters. The reduced Racah algebra $\tilde{\mathcal{R}}$ can be defined as the unital associative algebra generated by $R_{1}, R_{2}$ and $R_{3}$ with the following relations
\begin{align}\label{redRacah}
 [R_1,R_2] &= R_3,\nonumber \\
 [R_2,R_3] &= R_2^2 + \{R_1,R_2\} + d R_2 + e_{1},\\
 [R_3,R_1] &= R_1^2 + \{R_1,R_2\} + d R_1 + e_{2},\nonumber
\end{align}
where $d$ and $e_{i}$ for $i=1,2$ are central elements. The associated Casimir element \eqref{racah-casimir} is given by
\begin{align*}
C = \{ R_{1}^{2} , R_{2} \} + \{ R_{1}, R_{2}^{2} \} + R_{1}^{2} + R_{2}^{2} + R_{3}^{2} + (d+1) \{ R_{1}, R_{2} \} + (2 e_{1}+d) R_{1} + (2 e_{2} +d) R_{2}.
\end{align*}
Provided the parameters $a_1$ and $a_2$ in the Racah algebra \eqref{racah-rel} are non-vanishing, this reduced form is arrived at under the following affine transformation of the generators
\begin{align}\label{to-red-racah}
 K_1 &\mapsto a_2 R_1 - \frac{c_2}{2 a_2}I, \quad K_2 \mapsto a_1 R_2 -\frac{c_1}{2 a_1}I, \quad K_3 \mapsto a_1 a_2 R_3,
\end{align}
where
\begin{align*}
d &= \frac{a_2 a_1 b - a_1^2 c_2-a_2^2 c_1}{a_1^2 a_2^2}, \qquad e_{1} = \frac{-2 a_1 c_1 b+a_2 c_1^2+4 a_1^2 d_1}{4 a_1^4 a_2}, \qquad e_{2} = \frac{-2 a_2 b c_2+a_1 c_2^2+4 a_2^2 d_2}{4 a_1 a_2^4}.
\end{align*}
 
From the presentation \eqref{redRacah} of the (reduced) Racah algebra, one obtains the equitable presentation by introducing four generators as follows
\begin{align*}
V_{1} &= -2 R_{1}, \qquad V_{2} = -2 R_{2}, \qquad V_{3} = 2(R_{1}+R_{2}+d), \qquad P = 2 R_{3},
\end{align*}
such that one has that
\begin{align}\label{eq-racah-1}
V_{1}+V_{2}+V_{3} &= 2 d, \quad \text{and} \quad [V_{1},V_{2}]=[V_{2},V_{3}]=[V_{3},V_{1}] = 2 P.
\end{align}
The relations \eqref{redRacah} can be written in terms of the new generators as
\begin{align}\label{eq-racah-2}
[V_{1},P] &= V_{2}V_{1}-V_{1}V_{3}+4 e_{2}, \quad [V_{2},P] = V_{3}V_{2}-V_{2}V_{1}-4e_{1}, \quad [V_{3},P] = V_{1}V_{3}-V_{3}V_{2} + 4(e_{1}-e_{2}).
\end{align}
The presentation of the (reduced) Racah algebra given by \eqref{eq-racah-1} and \eqref{eq-racah-2} is referred to \cite{Genest2015} as the equitable presentation, as it makes manifest the $\mathbb{Z}_{3}$ symmetry of the Racah algebra given by the cyclic permutations of the generators. One concludes that for non-vanishing $a_{1}$ and $a_{2}$ there is an isomorphism $\chi: \mathcal{R} \longrightarrow \tilde{\mathcal{R}}$ that identifies the equitable presentation given by \eqref{eq-racah-1} and \eqref{eq-racah-2} with the Racah algebra as presented in \eqref{racah-rel}. Explicitly, this map is
\begin{align}\label{to-eq-racah}
\chi:\quad K_{1} \mapsto -	\frac{a_{2}}{2} V_{1} - \frac{c_2}{2 a_2}I, \quad K_2 \mapsto -\frac{a_1}{2} V_2 -\frac{c_1}{2 a_1}I, \quad K_3 \mapsto \frac{a_1 a_2}{2} P.
\end{align}

\subsection{Difference operator realization}
The Racah grid $\lambda$ is a two-parameter quadratic grid indexed by $x$ given by
\begin{align}\label{racah-grid}
 \lambda(x) = x(x+\gamma+\delta+1), \qquad x=0,1,\dots,N, \qquad N\in\mathbb{N}
\end{align}
where $\gamma$ and $\delta$ are real parameters. Defining the shift operators acting on functions of $x$ as
\begin{align}\label{shift-op}
 T^+ f(x) \mapsto f(x+1), \quad T^- f(x) \mapsto f(x-1),
\end{align}
one introduces the forward and backward difference operators as
\begin{align}\label{fb-op}
 \Delta = T^{+} - I, \quad \nabla = I - T^{-}.
\end{align}
With these notations, the Racah operator $Y$ takes the following form
\begin{align}\label{racah-op}
Y = B(x)\up - D(x) \dn,
\end{align}
where
\begin{align}\label{racah-real-coef}
 B(x) &= \frac{(x+\alpha+1)(x+\beta+\delta+1)(x+\gamma+1)(x+\gamma+\delta+1)}{(\dy+1)(\uy)},\\
 D(x) &= \frac{x(x-\alpha+\gamma+\delta)(x-\beta+\gamma)(x+\delta)}{(\dy)(\dy+1)}.
\end{align}
This operator is diagonalized \cite{Koekoek2010} by the four-parameters Racah polynomials $R_{n}(\lambda(x);\alpha,\beta,\gamma,\delta)$ defined for $n=0,1,2,\dots,N$ with $N\in \mathbb{N}$ and where either $$\alpha+1=-N, \qquad \beta+\delta+1=-N \quad\text{or}\quad \gamma+1=-N.$$
The eigenvalue equation is then
\begin{align*}
\left( B(x)\up - D(x) \dn \right) \, R_{n}(\lambda(x);\alpha,\beta,\gamma,\delta) = n(n+\alpha+\beta+1) R_{n}(\lambda(x);\alpha,\beta,\gamma,\delta).
\end{align*}
These Racah polynomials being orthogonal polynomials in the Askey scheme, they also satisfy \cite{Koekoek2010} a three-term recurrence relation of the following form
\begin{align*}
\lambda(x) R_{n}(\lambda(x)) = A_{n} R_{n+1}(\lambda(x))-(A_{n}+C_{n}) R_{n}(\lambda(x))+C_{n} R_{n-1}(\lambda(x)),
\end{align*}
where the coefficients $A_{n}$ and $C_{n}$ only depend \cite{Koekoek2010} on the parameters $\alpha, \beta, \gamma, \delta$ and the degree $n$. The left-hand side of the above can be understood as an operator that acts by multiplication on functions on $\lambda$. Denoting this recurrence operator as $X$
\begin{align*}
X=\lambda(x),
\end{align*}
a realization of the Racah algebra \eqref{racah-rel} in terms of second-order difference operators is given by
\begin{align}\label{racah-diff}
K_{1} \longmapsto Y, \qquad K_{2} \longmapsto X.
\end{align}
In this realization, the central elements in the relations \eqref{racah-rel} are proportional to the identity element. The scaling of these central elements and the parameters are determined by the parameters of the associated Racah operator as follows
\begin{align*}
a_1&=-2, & c_1 &= -(\gamma +\delta ) (\gamma +\delta +2), & d_1 &= -(\alpha +1) (\gamma +1) (\beta +\delta +1) (\gamma +\delta ) I,\\
a_2&=-2, & c_2 &= -(\alpha+\beta ) (\alpha +\beta +2), & d_2 &= -(\alpha +1) (\gamma +1) (\alpha +\beta ) (\beta +\delta +1) I,
\end{align*}
\begin{align*}
b &= 2\left[\beta(\delta-\alpha) - (\alpha+\beta) (\gamma +\delta +2)-2 (\gamma +1) (\delta +1)\right] I,\\
C &=(\alpha +1) (\gamma +1) (\beta +\delta +1) \big[2 \beta\delta -2\alpha + \beta(\alpha+1)(\gamma -1) + (\alpha-1)(\gamma +1) (\delta +1)\big] I.
\end{align*}
With these observations, the bispectral problem associated to the Racah algebra is completely specified. Moreover, the Racah polynomials are seen to span finite-dimensional representations of the Racah algebra under the realization \eqref{racah-diff}.

\section{The Heun-Racah operator}\label{heun-racah-operator}

This section is concerned with the construction of a generalization of the Heun differential operator on the Racah grid \eqref{racah-grid}. The key property of the standard differential Heun operator is that it is the most general second-order differential operator that sends polynomials of degree $n$ to polynomials of degree $n+1$. By requiring an equivalent property for operators on the Racah grid, one obtains the desired generalization.

Consider the vector space $\mathcal{P}$ of polynomials on the Racah grid $\lambda(x)$ as given by \eqref{racah-grid}. One then considers difference operators expressed in terms of the shift operators \eqref{shift-op}. In this setting, a generic second-order difference operator on $\mathcal{P}$ can be written as
\begin{align}\label{wfirstdef}
 W = A_1(x) T^{+} + A_2(x) T^{-} + A_0(x) I.
\end{align}
With the forward $\Delta$ and backward $\nabla$ difference operators defined in \eqref{fb-op}, one can write \eqref{wfirstdef} as the second-order difference operator
\begin{align}\label{wdef}
 W = A_1(x) \up - A_2(x) \dn + \left[ A_1(x) + A_2(x) + A_0(x) \right] I.
\end{align}
The Heun-Racah operator is now defined as the most general degree increasing second-order difference operator $W$ such that
\begin{align}\label{heuncondition}
 W:\pol \rightarrow \pol : p_n(\lambda) \mapsto q_{n+1}(\lambda),
\end{align}
for $p_n$ and $q_{n+1}$ arbitrary polynomials in $\pol$ of degree $n$ and $n+1$, respectively.

\subsection{Parametrization of the Heun-Racah operator}
The condition \eqref{heuncondition} determines the form of the functions $A_i(x)$ for $i=0,1,2$ in \eqref{wdef}. This can be seen by acting with $W$ on monomials in $\pol$ and demanding that \eqref{heuncondition} holds. Observing that
\begin{align*}
 \Delta \cdot \y = \uy, \qquad \nabla \cdot \y = \dy,
\end{align*}
one has
\begin{align}\label{Aconstraints}
\left\{
\begin{aligned}
  W \cdot 1 &=A_0(x) + A_1(x) + A_2(x) = p_1(\lambda(x))\\
  W \cdot \lambda(x) &= (\y + \uy) A_1(x) + (\y - (\dy)) A_2(x) + \y A_0(x) = p_2(\y)\\
  W \cdot \y^2 &= (\y + \uy)^2 A_1(x) + (\y - (\dy))^2 A_2(x) + \y^2 A_0(x) = p_3(\y),
\end{aligned}\right.
\end{align}
with $p_{1},p_{2}$ and $p_{3}$ being arbitrary polynomials of first, second and third degree, respectively. From \eqref{Aconstraints}, one finds that
\begin{align*}
 A_0(x) &= p_1(\lambda(x)) - A_2(x) - A_1(x),\\
 \al(x)(A_1(x)-A_2(x)) &= p_2(\lambda(x)) - \lambda(x) p_1(\lambda(x)) - 2 A_1(x),\\
 \al^2(x) (A_1(x) + A_2(x)) &= p_3(\lambda(x)) - 2\lambda(x) p_2(\lambda(x)) + \lambda^2(x) p_1(\lambda(x)) - 4(\al(x)+1) A_1(x),
\end{align*}
where $\al(x) = \nabla \lambda(x) = \dy$. Introducing the polynomials $\pi_{1}$, $\pi_{2}$ and $\pi_{3}$
\begin{equation}\label{pidef}
\begin{aligned}
 \pi_{1}(\y) &= p_{1}(\y), \qquad \pi_2(\y) = \frac{p_2(\y) - \y p_1(\y)}{2},\\
 \pi_3(\y) &= \frac{p_3(\y) - 2\y p_2(\y) + \y^2 p_1(\y)}{2},
 \end{aligned}
\end{equation}
parametrized as follows
\begin{align}\label{heun-racah-op-param}
 \pi_1(z) &= \sum_{i=0}^{1} t_i z^i, & \pi_2(z) &= \sum_{i=0}^2 u_i z^i, & \pi_3(z) &= \sum_{i=0}^3 v_i z^i,
\end{align}
one solves for the $A_i(x)$ to get
\begin{equation}\label{Adef}
\begin{aligned}
 A_1(x) &= \frac{\pi_3(\y) + \al(x) \pi_2(\y)}{(\al(x) +1)(\al(x) + 2)}, \qquad A_2(x) = \frac{\pi_3(\y) - (\al(x)+2)\pi_2(\y)}{ \al(x)(\al(x)+1)},\\
 A_0(x) &= \pi_1(\y) - A_1(x) - A_2(x).
 \end{aligned}
\end{equation}
It follows from \eqref{Adef} that specifying the coefficients of the polynomials $\pi_1$, $\pi_2$ and $\pi_3$ fully determines the functions $A_{i}(x)$, $i=0,1,2$ in \eqref{wdef} upon fixing the the grid $\lambda(x)$. Together, these polynomials admit nine free parameters. However, in order for the Heun-Racah operator to act on the finite grid $\y$, one must have that $(x-N)$ is a factor of $A_{1}(x)$ and $x$ is a factor of $A_{2}(x)$. The second condition is uniquely satisfied by demanding that
\begin{align*}
v_{0}=u_{0}(\gamma+\delta+2).
\end{align*}
To satisfy the first one, one finds that the following must hold:
\begin{multline*}
v_{1}=-\frac{2 u_{0}}{N} - (N+\gamma+\delta+1)^2 N^2 v_3 - (N+\gamma+\delta+1)N v_2\\
- \big[2 N(N+1) + (3N+\gamma+\delta+1)(\gamma+\delta)\big]N u_2 - (2 N+\gamma+\delta) u_1.
\end{multline*}
Thus, with these constraints and a fixed grid $\y$, the Heun-Racah operator admits seven free parameters. In the parametrization \eqref{heun-racah-op-param} of \eqref{Adef}, the remaining parameters are $t_{0}, t_{1}, u_{0}, u_{1}, u_{2}, v_{2}$ and $v_{3}$.

\subsection{Sufficiency of the construction}
It remains to show that the operator $W$ specified by \eqref{wdef} and \eqref{Adef} satisfies the property \eqref{heuncondition} in general. To do so, one first computes the action of $W$ on a generic monomial $ \y^{n} \in\pol$ to obtain
\begin{align}\label{gen-mon}
W \cdot \y^{n} = (\y + \al(x)+2)^{n} A_{1}(x) + (\y - \al(x))^{n} A_{2}(x) + \y^{n} A_{0}(x).
\end{align}
Expanding the binomials, the above can be written as
\begin{align}
W \cdot \y^{n} &= \sum\limits_{k=1}^{n} \bin{n}{k} \y^{n-k} \chi_{k} + \y^{n}\, \pi_{1}(\y),\label{gen-mon-exp}\\
\chi_{k} &= \big[ (\al(x)+2)^{k} A_{1}(x) + (-\al(x))^{k} A_{2}(x) \big]\label{pterm},
\end{align}
where the last term in \eqref{gen-mon-exp} is manifestly a polynomial in $\y$ of degree $n+1$ as $\pi_{1}$ is a linear function by construction. Using binomial expansions, one has that
\begin{align*}
(-\al(x))^{k-1} = \sum\limits_{j=0}^{k-1} \bin{k-1}{j} (-1)^{j} (\al(x)+1)^{j}, \qquad (\al(x)+2)^{k-1} = \sum\limits_{j=0}^{k-1} \bin{k-1}{j} (\al(x)+1)^{j}.
\end{align*}
The above identities with \eqref{Adef} in \eqref{pterm}, leads to
\begin{align*}
\chi_{k} &= \sum\limits_{j=0}^{k-1} \bin{k-1}{j} (\al(x)+1)^{j-1} \bigg[ \big[\pi_{3}(\y)+\al(x) \pi_{2}(\y)\big]+(-1)^{j-1}\big[\pi_{3}(\y)-(\al(x)+2)\pi_{2}(\y)\big] \bigg],\\
 &= \sum\limits_{\substack{j\text{ even}\\ 0\leq j \leq k-1 }} \bin{k-1}{j} 2 \pi_{2}(\y) (\al(x)+1)^{j} + \sum\limits_{\substack{j\text{ odd}\\ 0\leq j \leq k-1 }} \bin{k-1}{j} 2 \big[\pi_{3}(\y) - \pi_{2}(\y)\big] (\al(x)+1)^{j-1}.
\end{align*}
It is readily verified that $(\al(x)+1)^{2} = 4\y + (c+d+1)^{2}$ is a linear function of $\y$ such that even powers of $(\al(x)+1)$ can be identified as polynomials in $\y$. Thus, it can be seen that $\chi_{k}$ is a polynomial in $\y$ of degree
\begin{align*}
\mathrm{deg}(\chi_{k}) =
\begin{cases}
\frac{k}{2}+2 & \text{for }k\text{ even,}\\
\frac{k+3}{2} & \text{for }k\text{ odd.}
\end{cases}
\end{align*}
In particular, the degree of $\chi_{k}$ is less than $k+1$ for $k>2$. Thus, one can conclude from the above and \eqref{gen-mon-exp} that the operator $W$ acts on monomials as follows
\begin{align}\label{gen-mon-sol}
W \cdot \y^{n} = (t_{1}+ 2n\, u_{2} + n(n-1)\, v_{3}) \y^{n+1} + O(\y^{n}),
\end{align}
where $t_{1},\, u_{2}$ and $v_{3}$ are parameters of the Heun-Racah operator as labelled in \eqref{heun-racah-op-param}. This result implies that $W$ as defined by \eqref{wdef} and \eqref{Adef} is the most generic second order difference operator on the grid $\y$ that satisfies property \eqref{heuncondition} provided that
\begin{align*}
t_{1} \neq 0, \qquad u_{2}\neq 0,\quad\text{or}\quad v_{3} \neq 0.
\end{align*}

\subsection{Specialization as the Racah operator}
The Heun-Racah operator was constructed as the most general second-order operator on the Racah grid that satisfies the degree raising property \eqref{heuncondition}. We now consider specializations that preserve the space of polynomials of degree $n$ in $\mathcal{P}$. From \eqref{gen-mon-sol}, one easily identifies the necessary constraints to be
\begin{align}\label{p-stab-cons}
t_{1}=u_{2}=v_{3}=0.
\end{align}
Furthermore, if the above constraints are satisfied, normalizing the Heun-Racah operator so that the numerators of the functions $A_{1}(x)$ and $A_{2}(x)$ are monic polynomials corresponds to setting
\begin{align}\label{monic-cons}
v_{2}=1.
\end{align}
Finally, from \eqref{gen-mon-exp}, one can identify the term of $W$ proportional to the identity as $t_{0} I$. This term vanishes if
\begin{align}\label{no-id-cons}
t_{0}=0.
\end{align}
Upon demanding that \eqref{p-stab-cons}, \eqref{monic-cons} and \eqref{no-id-cons} are satisfied, the Heun-Racah operator \eqref{wdef} takes the form of the Racah operator. The remaining free parameters in \eqref{heun-racah-op-param} can be expressed in terms of those of the Racah operator on the same grid $\y$ of size $N$ as follows
\begin{align*}
u_{0}=\frac{(\alpha +1) (\gamma +1) (\beta +\delta +1)}{2}, \quad u_{1}= \frac{\alpha+\beta+2}{2}.
\end{align*}

\subsection{Relation with the algebraic Heun operator}
The Heun-Racah operator $W$ given by \eqref{wdef} together with \eqref{Adef} can be written as a bilinear combination of the Racah algebra generators in the canonical realization \eqref{racah-diff} in terms of difference operators. This bilinear expression is shown to coincides with the canonical construction of the algebraic Heun operator as follows. Recall, that the Racah generators in this realization are given by
\begin{align*}
 K_2 = X = \lambda(x), \qquad K_1 = Y = B(x)\up - D(x) \dn,
\end{align*}
where $\lambda(x)$ is the Racah grid \eqref{racah-grid} of size $N$ and the coefficients $B(x)$ and $D(x)$ are given by \eqref{racah-real-coef}. Consider the operator
\begin{align}\label{racah-op-from-real}
 W = \tau_1 XY + \tau_2 YX + \tau_3 X + \tau_4 Y + \tau_0 I,
\end{align}
as in the introduction. A direct computation leads to
\begin{multline}\label{w-from-tridig}
 W = \Big[\y (\tau_1+\tau_2) + \tau_4 + \tau_2 (\al(x) +2) \Big]B(x) \up\\
 - \Big[\y ( \tau_1 + \tau_2) + \tau_4 - \tau_2 \al(x) \Big] D(x) \dn\\
 + \tau_2 \Big[ (\al(x)+2)B(x) - \al(x) D(x) \Big] + \tau_3 \y + \tau_0,
\end{multline}
where, again, $\al(x) = \dy$. Equation \eqref{w-from-tridig} defines an operator of the form \eqref{wdef} with 
\begin{align}
A_{1}(x) &= \Big[\y (\tau_1+\tau_2) + \tau_4 + \tau_2 (\al(x) +2) \Big]B(x), \qquad A_{2}(x) = \Big[\y ( \tau_1 + \tau_2) + \tau_4 - \tau_2 \al(x) \Big] D(x),\nonumber\\
A_{0}(x) &= \tau_3 \y + \tau_0-\Big[\y (\tau_1+\tau_2) + \tau_4\Big]\big(B(x)+D(x)\big) .
\end{align}
Using \eqref{Adef}, one can express the polynomials $\pi_{1}$, $\pi_{2}$ and $\pi_{3}$ in terms of $A_{i}(x)$, $i=0,1,2$ and $\al(x)$ as follows
\begin{equation}\label{pi-from-a}
\begin{aligned}
 \pi_{1}(\y) &= A_{0}(x) + A_{1}(x) + A_{2}(x), \qquad \pi_3(\y) = \frac{(\al(x)+2)^{2} A_1(x) + \al^{2}(x) A_{2}(x) }{2},\\
 \pi_2(\y) &= \frac{(\al(x) + 2) A_1(x) - \al(x) A_2(x)}{2}.
 \end{aligned}
\end{equation}
The above allows one to relate the operator \eqref{w-from-tridig} with the Heun-Racah operator. Indeed, comparing the terms, one finds that \eqref{w-from-tridig} can be identified with the Heun-Racah operator defined by \eqref{wdef} and \eqref{Adef} provided the parameters \eqref{heun-racah-op-param} are given by
\begin{align}
 u_0 &= \left(\tau
   _2 (\gamma +\delta +2)+\tau _4\right)\phi_{\alpha,\beta,\gamma,\delta} & t_0 &= 2 \tau_2\phi_{\alpha,\beta,\gamma,\delta}+\tau_0,\nonumber\\
 u_1 &= (\tau _1+\tau_2) \phi_{\alpha,\beta,\gamma,\delta} + \tau _2 \psi_{\alpha,\beta,\gamma,\delta}+\tau _4 (\alpha +\beta +2)/2 & t_1 &= \tau_2(2+\alpha+\beta)+\tau_3,\nonumber\\
 u_2 &= (\tau _1+\tau_2) (\alpha +\beta +2)/2 + \tau _2 , & v_3 &= \tau _1+\tau _2,\nonumber\\
 v_2 &= (\tau _1+\tau_2) \psi_{\alpha,\beta,\gamma,\delta} + 2 \tau_2 (\alpha+\beta+3)+\tau_4, &&\label{aho-to-heun-racah}
\end{align}
where 
\begin{align*}
\phi_{\alpha,\beta,\gamma,\delta} &= (\alpha +1) (\gamma +1) (\beta +\delta +1)/2,\\
\psi_{\alpha,\beta,\gamma,\delta} &= \alpha  \left(\beta +\frac{\gamma +\delta}{2}
   +2\right)+\beta \left(\frac{\gamma -\delta}{2} +2\right)+ (\gamma  \delta +\gamma
   +\delta +3).
\end{align*}
We remind the reader that the realization \eqref{racah-diff} of the Racah algebra admits two free parameters, once the grid $\y$ of size $N$ is specified. Moreover, the construction \eqref{racah-op-from-real} for the algebraic Heun operator introduces five additional parameters. Thus, the seven free parameters in \eqref{heun-racah-op-param} of the Heun-Racah operator \eqref{wdef} are in correspondence with those of the algebraic Heun operator of the Racah type in the canonical realization \eqref{racah-diff}


\section{The Heun-Racah algebra}\label{heun-racah-algebra}
The Heun-Racah algebra $\mathcal{HR}$ is introduced as the unital associative algebra over $\C$ generated by $X,W,Z$ with the following relations
\begin{equation}\label{hralgebra}
\begin{aligned} 
{}[W,X] &=Z,\\
[X,Z] &= x_{0} + x_{1} X + x_{2} X^{2} + x_{3} X^{3} + x_{4} W + x_{5} \{ X, W \},\\
[Z,W] &= y_{0} + y_{1} X +y_{2} X^{2} + y_{3} X^{3} + (x_{1}-x_{3}x_{4}) W + x_{5} W^{2} + (x_{2}-x_{3}x_{5})\{X,W\} + 3 x_{3} XWX,
\end{aligned}
\end{equation}
where $x_{i}\in\R$ for $i=3,4,5$ and $y_{3}\in\R$ are free parameters and where $x_{i}$, $y_{i}$ for $i=0,1,2$ are central elements. The constraints on the last three coefficients in \eqref{hralgebra} ensure compatibility with the following Jacobi identity
\begin{align*}
[[X,Z],W]+[[Z,W],X]+[[W,X],Z] = 0.
\end{align*}
One readily notices that the relations \eqref{hralgebra} reduce to the relations \eqref{racah-rel} of the Racah algebra if
\begin{align}\label{collapse-to-racah}
x_{3}=y_{2}=y_{3}=0, \qquad x_{2}\propto I \qquad \text{and} \qquad y_{1}\propto I.
\end{align}
It is verified that the element $\Omega \in \mathcal{HR}$ given by
\begin{align}\label{heun-racah-casimir}
\Omega = e_{1} X + e_{2} W + e_{3} \{X,W\} + e_{4} XWX + e_{5} WXW + e_{6} X^{2} + e_{7} W^{2} - Z^{2} + [XW,WX] + e_{8} X^{3} + e_{9} X^{4},
\end{align}
is central when the coefficients are as follows
\begin{align}
e_{1} &=x_5 y_1+x_4 y_2/3+x_4 x_5 y_3/6-y_0, & e_{4} &=4 x_3 x_5-x_2, & e_{7} &=-2 x_4,\\
e_{2} &=x_2 x_4-3 x_0- x_3 x_4 x_5, & e_{5} &=-3 x_5, & e_{8} &=(5 x_5 y_3+y_2)/3,\\
e_{3} &=x_3 x_4+ x_2 x_5- x_3 x_5^2- x_1, & e_{6} &=x_5^2 y_3/6+4 x_5 y_2/3+x_4 y_3/2, & e_{9} &=y_3/2.
\end{align}

\subsection{Embedding in the Racah algebra}
An embedding of a specialization of the Heun-Racah algebra $\mathcal{HR}$ in the Racah algebra $\mathcal{R}$ is possible. This specialization is obtained by imposing conditions on the parameters of the Heun-Racah relations \eqref{hralgebra}. Consider the mapping defined by
\begin{align}\nonumber
\Phi : \mathcal{HR} &\longrightarrow \mathcal{R},\\
X &\longmapsto K_{2},\label{phi-emb}\\
W &\longmapsto \tau_1 K_{2} K_{1}+ \tau_2 K_{1} K_{2} + \tau_3 K_{2} + \tau_4 K_{1} + \tau_0 I.\nonumber
\end{align}
The mapping $\Phi:\mathcal{HR}\rightarrow \mathcal{R}$ is an algebra homomorphism provided, first, that the parameters of the Heun-Racah algebra be the following functions of the parameters of the Racah
\begin{align}
x_{3} &= a_2 (\tau _1+\tau _2), \qquad x_{4} = c_{1}, \qquad x_{5} = a_{1},\nonumber\\
y_{3} &= 2 a_2^2\, \tau_1 \tau_2 - 4 a_2 \tau_{3} \left(\tau _1+\tau _2\right)+2 c_2 \left(\tau
   _1+\tau _2\right)^2,\label{special1}
\end{align}
and, second, that the central elements be mapped to those of the Racah algebra as follows
\begin{multline*}
x_{0} \mapsto \tau_4 d_1 - c_1 \tau_0, \qquad x_{1} \mapsto (\tau _1+\tau _2) d_1 + \tau_4 b - 2 a_1 \tau_0 - c_1 \tau_3, \qquad x_{2} \mapsto b (\tau _1+\tau _2) + \tau_4 a_2 - 2 a_1 \tau_3,
\end{multline*}
\begin{multline*}
y_{0} \mapsto \left[ a_1 C+b d_1+(a_1^2-c_1) d_2\right] \tau_{1} \tau_{2} + \Big[ \left(a_2 c_1-d_1\right)\tau_0 - \left(C+a_2 d_1+a_1 d_2\right)\tau_4\Big] (\tau_{1}+\tau_{2})\\
+a_1 \tau _0^2+(d_2 \tau _4-b\tau _0) \tau _4+(c_1 \tau _0-d_1 \tau _4)\tau _3,
\end{multline*}
\begin{multline*}
y_{1} \mapsto \big[b^2+a_1^2 c_2+2 a_2 d_1-c_1 c_2-a_1 (a_2 b+4 d_2)\big] \tau_{1} \tau_{2} - \big[C+a_2 d_1+a_1 d_2\big](\tau_{1}+\tau_{2})^{2}+\left(4 a_1 \tau _0-2 b \tau _4 + c_{1}\tau_{3} \right)\tau _3\\
+\big[ \left(2 a_1 a_2-2 b\right)\tau _0 + \left(4 d_2-a_1 c_2\right)\tau _4 + (a_2 c_1-2 d_1) \tau _3\big](\tau_{1}+\tau_{2})+(c_2 \tau_4-2 a_2 \tau_0)\tau_4,
\end{multline*}
\begin{multline}
y_{2} \mapsto \big[3 d_2-a_1 c_2\big] (\tau
   _1+\tau _2)^2 + \big[3 a_2 b-3 a_1 c_2-a_1 a_2^2\big] \tau_1 \tau_2 + \big[ \left(2 a_1 a_2-3 b\right)\tau_3 - 3 a_2 \tau _0+3 c_2 \tau _4\big](\tau_1+\tau_2)\\
 + 3 (a_1 \tau_3 - a_2 \tau_4)\tau_3,\qquad \label{special2}
\end{multline}
The above specialization of the Heun-Racah algebra admits a realization in terms of the Heun-Racah operator \eqref{wdef}. Indeed, comparing \eqref{phi-emb} with \eqref{racah-op-from-real}, one can see that the generator $W$ of this specialized Heun-Racah algebra is embedded in the Racah algebra as the algebraic Heun operator of the Racah type. A natural realization of the specialization \eqref{special1} and \eqref{special2} of the Heun-Racah algebra is obtained from the concatenation of the embedding \eqref{phi-emb} and the canonical realization \eqref{racah-diff} of the Racah algebra. In this realization, one finds that $W$ takes the form of \eqref{w-from-tridig}, which was identified as the Heun-Racah operator with parameters given by \eqref{aho-to-heun-racah}. Moreover, it follows from the above that in the specialization \eqref{collapse-to-racah} of the Heun-Racah algebra to the Racah algebra, the map \eqref{phi-emb} is an affine transformation of the Racah algebra parametrized by $\tau_{0}$ and $\tau_{4}$.

With the parameters of the Heun-Racah algebra as in \eqref{special1}, one finds that the image of the central element $\Omega$ given in \eqref{heun-racah-casimir} under the mapping $\Phi$ given by \eqref{phi-emb} and \eqref{special2} is the Casimir element \eqref{racah-casimir} of the Racah algebra, up to a central element and scaling. Explicitly, one has
\begin{align*}
\Phi: \Omega \longmapsto u\, C + v,
\end{align*}
where the coefficients $u$ and $v$ are given by
\begin{align*}
u = \left[(c_{1}-a_1^2)  \tau _1 \tau _2 +a_1 \left(\tau _1+\tau _2\right) \tau
   _4 - \tau _4^2\right]^{-1},
\end{align*}
and
\begin{multline*}
v= u \Big[\tau _1 \tau _2 (a_1 b d_1-a_1 c_1 d_2-a_2 c_1 d_1+a_2 a_1^2 d_1-d_1^2)+ \big[(a_1 a_2 c_1-b c_1)\tau_0 + (c_1 d_2-2 a_1 a_2 d_1)\tau_4 \big](\tau _1+\tau _2)\\
   +(2 a_1 c_1 \tau_0-2 a_1 d_1 \tau_4)\tau_{3}+\big(2a_{2}d_{1}\tau_{4} + (2d_{1}-a_{2}c_{1})\tau_{0}\big)\tau_{4} -c_{1} \tau_{0}^{2}  \Big] + a_{2} d_{1} - a_{1} d_{2}.
\end{multline*}

\section{The Bannai-Ito algebra}\label{bannai-ito}
The Bannai-Ito algebra $\mathcal{B}$ is defined \cite{DeBie2015} as the unital associative algebra over $\C$ generated by $B_1,B_2$ and $B_3$ with the following relations
\begin{align}\label{bi-algebra}
\{ B_1, B_2 \} = B_3 + \omega_{1}, \quad \{B_2,B_3\}= B_1 + \omega_{2}, \quad \{B_1,B_3\}=B_2+\omega_{3},
\end{align}
where $\omega_{i}$ for $i=1,2,3$ are central elements. A natural $\mathbb{Z}_{2}$ grading is given by taking $B_1,B_2$ to be odd, which implies that $B_3$ is even. It is observed that this algebra satisfies the following graded Jacobi identity
\begin{align*}
[B_1,\{B_2,B_3\}]+[B_2,\{B_1,B_3\}]+[B_3,\{B_1,B_2\}]=0.
\end{align*}
In this presentation, the central Casimir operator is given by
\begin{align}\label{bi-casimir}
Q = B_1^{2}+B_2^{2}+B_3^{2}.
\end{align}

\subsection{Canonical realization}
A realization of the Bannai-Ito algebra in terms of reflection operators acting on univariate polynomials can be constructed \cite{DeBie2015} as follows. One first defines two reflection operators $R_{1}$ and $R_{2}$ acting on univariate functions as
\begin{align*}
R_{1}\, f(x) = f(-x), \qquad R_{2}\, f(x) = f(-x-1), \quad \implies \quad R_{1}^{2}=R_{2}^{2}=I.
\end{align*}
The most general symmetrizable first order shift operator that contains reflections and preserves the space of polynomials of a given degree is the Bannai-Ito operator \cite{Tsujimoto2012} given by
\begin{align}\label{bi-op}
\tilde{B_2}= \frac{\left(x-\rho _1\right) \left(x-\rho _2\right)}{2 x}(1-R_{1})+\frac{\left(-r_1+x+\frac{1}{2}\right) \left(-r_2+x+\frac{1}{2}\right)}{2 x+1}(R_{2}-1).
\end{align}
This operator is diagonalized by the four parameters Bannai-Ito polynomials \cite{Bannai1984,Vinet2010}, denoted $B_{n}(x\rvert\rho_{1},\rho_{2},r_{1},r_{2})$, with
\begin{align}\label{bi-op-action}
\tilde{B_2}\, B_{n}(x\rvert\rho_{1},\rho_{2},r_{1},r_{2}) = (-1)^{n}(n+\rho _1+\rho _2-r_1-r_2+1/2) B_{n}(x\rvert\rho_{1},\rho_{2},r_{1},r_{2}).
\end{align}
These polynomials are orthogonal on the finite Bannai-Ito grid $x_{s}$ defined \cite{Genest2013g}, depending on the truncation conditions used, as
\begin{align}\label{bi-grid}
x_{s} &=
\begin{cases}
(-1)^{s}\left( \frac{s}{2} + \rho_{j} + \frac{1}{4} \right) - \frac{1}{4}, & \text{for }N\text{ even, } 2(r_{i}+\rho_{j}) = N+1,\, i,j=1,2,\\
(-1)^{s}\left( \frac{s}{2} + \rho_{2} + \frac{1}{4} \right) - \frac{1}{4}, & \text{for }N\text{ odd, }2(\rho_{1}+\rho_{2}) = -N-1,\\
(-1)^{s}\left( r_{1} - \frac{s}{2} - \frac{1}{4} \right) - \frac{1}{4}, & \text{for }N\text{ odd, }2(r_{1}+r_{2}) = N+1.
\end{cases} & \text{with }s=0,1,\dots,N\, \in \mathbb{N},
\end{align}
The Bannai-ito polynomials also satisfy a three-terms recurrence relation of the following form
\begin{align}\label{bi-rec}
x B_{n}(x) = B_{n+1}(x)+A_{n} B_{n}(x)+C_{n} B_{n-1}(x),
\end{align}
where the coefficients depend only \cite{Tsujimoto2012} on $n$ and the parameters $\rho_{1},\rho_{2},r_{1},r_{2}$ of the polynomial. The left-hand side of this recurrence relation can be understood as an operator $\tilde{B_1}$ that acts on functions by multiplication as follows
\begin{align}\label{bi-rec-op}
\tilde{B_1} f(x) = x f(x).
\end{align}
As defined in \eqref{bi-op} and \eqref{bi-rec-op}, the pair of operators $\tilde{B_{1}}$, $\tilde{B_{2}}$ acts on univariate functions of $x$. Another realization can be given where both operators act on the degrees $n$. In this case, the action of $\tilde{B_{2}}$ is given in \eqref{bi-op-action} while the action of $\tilde{B_{1}}$ is defined through the right-hand side of \eqref{bi-rec}. Thus, the operators $\tilde{B_{1}}$, $\tilde{B_{2}}$ form a bispectral pair.

Introducing the structure operator as $\tilde{B_3}\equiv \{\tilde{B_1},\tilde{B_2}\}$, it can be seen that the algebra generated by $\tilde{B_1},\tilde{B_2}$ and $\tilde{B_3}$ is the algebra \eqref{bi-algebra}, up to an affine transformation. Explicitly, the following map
\begin{align}\label{bi-real}
B_1&\longmapsto 2 \tilde{B_1}+1/2,\qquad B_2\longmapsto 2 \tilde{B_2}+(\rho _1+\rho _2-r_1-r_2+1/2), \\
B_3 &\longmapsto 4\{\tilde{B_1},\tilde{B_2}\}+2\tilde{B_{2}}+4\left(\rho_{1}+\rho_{2}-r_{1}-r_{2}+1/2\right)\tilde{B_{1}} + \left(\rho_{1}+\rho_{2}-4\rho_{1}\rho_{2}-r_{1}-r_{2}+4r_{1}r_{2}+1/2\right),\nonumber
\end{align}
is an homomorphism. In this realization, the Casimir operator \eqref{bi-casimir}, together with the central elements in \eqref{bi-algebra}, are proportional to the identity element. One has
\begin{align*}
 Q \longmapsto 2\,(\rho_1^2+\rho_2^2+r_1^2+r_2^2-1/8)\, I,
\end{align*}
\begin{align*}
\omega_{1}\mapsto 4 \left(\rho _1 \rho _2-r_1 r_2\right)I, \qquad \omega_{2}\mapsto 2 \left(\rho _1^2+\rho _2^2-r_1^2-r_2^2\right) I, \qquad \omega_{3}\mapsto 4 \left(\rho _1 \rho _2+r_1 r_2\right)I.
\end{align*}

\subsection{Embedding of the Racah algebra in the Bannai-Ito algebra}
An embedding of the (reduced) Racah algebra \eqref{redRacah} into the Bannai-Ito algebra has been shown to exists \cite{Genest2015}. This embedding is constructed from quadratic combinations of the Bannai-Ito generators as follows. One defines the generators $A,B$ and $C$ as
\begin{align}\label{abc-def}
A &= \frac{1}{4} \left( B_{1}^{2} -B_{1} -\frac{3}{4} \right), \qquad B = \frac{1}{4} \left( B_{2}^{2} -B_{2} -\frac{3}{4} \right), \qquad C = \frac{1}{4} \left( B_{3}^{2} -B_{3} -\frac{3}{4} \right).
\end{align}
A direct computation shows that in the sub-algebra generated by $A,B$ and $C$ as defined above, the element
\begin{align}\label{gamma-def}
\Gamma \equiv B_{1} + B_{2} + B_{3} - 3/2,
\end{align}
is central, as is the sum of the generators since it can be expressed as
\begin{align*}
A + B + C = \frac{1}{4} (Q - \Gamma -15/4),
\end{align*}
where $Q$ is the Casimir operator \eqref{bi-casimir} of the Bannai-Ito algebra. The commutators between distinct generators are seen to be equal and define a fourth generator $P$ as follows
\begin{align}\label{p-def}
2 P \equiv [A,B] = [B,C] = [C,A].
\end{align}
From the relations \eqref{bi-algebra} of the Bannai-Ito algebra, one obtains the closure of the algebra generated by $A,B,C$ and $P$ as a quadratic algebra with the following relations:
\begin{align}
[A,P] &= BA-AC+\frac{1}{16}\frac{\omega_{3}-\omega_{1}}{2}\left(\frac{\omega_{3}+\omega_{1}}{2}-\Gamma\right),\nonumber\\
[B,P] &= CB-BA+\frac{1}{16}\frac{\omega_{1}-\omega_{2}}{2}\left(\frac{\omega_{1}+\omega_{2}}{2}-\Gamma\right),\label{r-in-bi}\\
[C,P] &= AC-CB+\frac{1}{16}\frac{\omega_{2}-\omega_{3}}{2}\left(\frac{\omega_{2}+\omega_{3}}{2}-\Gamma\right).\nonumber
\end{align}
The relations \eqref{gamma-def}, \eqref{p-def} and \eqref{r-in-bi} can be seen to be identical to the equitable presentation of the Racah algebra given in \eqref{eq-racah-1} and \eqref{eq-racah-2}. Thus, one can define the embedding $\theta:\tilde{\mathcal{R}} \longrightarrow \mathcal{B}$ of the reduced Racah algebra into the Bannai-Ito algebra as follows

\begin{align}
\theta:\quad V_{1} &\mapsto A, & d &\mapsto \frac{1}{8} (Q - \Gamma -15/4),\nonumber\\
 V_{2} &\mapsto B, & e_{1} &\mapsto \frac{1}{64}\frac{\omega_{3}-\omega_{1}}{2}\left(\frac{\omega_{3}+\omega_{1}}{2}-\Gamma\right),\label{r-to-bi} \\
 V_{3} &\mapsto C, & e_{2} &\mapsto \frac{1}{64}\frac{\omega_{1}-\omega_{2}}{2}\left(\frac{\omega_{1}+\omega_{2}}{2}-\Gamma\right).\nonumber
\end{align}

\section{The Heun-Bannai-Ito algebra}\label{heun-bannai-ito}
We now introduce the Heun Bannai-Ito algebra $\mathcal{HB}$ abstractly as the unital associative cubic algebra generated by $X,W$ and $Z$ with the following relations
\begin{align}
 \{X,W\} &\equiv Z,\nonumber\\
 \{Z,X\} &= x_0 I + x_1 X + x_2 X^2 + x_3 X^3 + x_4 W,\label{hbi-alg}\\
 \{W,Z\} &= y_0 I + y_1 X + y_2 X^2 + y_3 X^3 + (x_1 + x_3 x_4) W + x_2 \{X,W\} - x_3 XWX\nonumber,
\end{align}
where $x_{i},\, y_{i}$ for $i=0,1,2$ are central elements and $x_{i}$ for $i=3,4$, together with $y_{3}$, are parameters in $\R$. The constraints in the last three coefficients of \eqref{hbi-alg} ensure compatibility with the following graded Jacobi identity
\begin{align*}
 [X,\{Z,W\}] + [W,\{X,Z\}] + [Z,\{W,X\}] = 0.
\end{align*}
A distinguished central element is identified in this presentation as
\begin{multline}\label{hbi-casimir}
 \Lambda= (x_4 y_2-y_0) X + (x_0-x_2 x_4)W - (x_1 + x_3 x_4)Z +\frac{1}{2}x_4 y_3 X^2 + 2 x_4 W^2 + Z^2 + [XW,WX]\\
 - x_2 XWX -y_2 X^3 -\frac{y_3}{2} X^4.
\end{multline}

\subsection{Embedding in the Bannai-Ito algebra}
A specialization of the Heun-Bannai-Ito algebra, obtained by imposing conditions on the parameters, admits an embedding into the Bannai-Ito algebra. This embedding can be constructed from the algebraic Heun operator \eqref{wdefalg} of the Bannai-Ito type. To do so, one defines the map $\psi$ on the generators as follows
\begin{align}
\psi:\mathcal{HB} &\longrightarrow \mathcal{B},\nonumber\\
 X &\longmapsto B_1,\nonumber\\
 W &\longmapsto \tau_1 B_1 B_2 + \tau_2 B_2 B_1 + \tau_3 B_{1} + \tau_4 B_{2} + \tau_0 I,\label{psi-map}
\end{align}
which is an homomorphism, provided that the parameters of the Heun-Bannai-Ito algebra in \eqref{hbi-alg} be as follows
\begin{align}\label{special1bi}
 x_3 &= 4 \tau_3, & x_4 &= 1, & y_3 &= 8 \tau _3^2-2 \left(\tau _1-\tau _2\right){}^2,
\end{align}
and that the central elements of the Heun-Bannai-Ito algebra be mapped to those of the Bannai-Ito algebra as follows
\begin{multline*}
 x_0 \mapsto \tau _4 \omega _3-\tau _0, \qquad\qquad x_1 \mapsto 2 \tau _4 \omega _1+\left(\tau _1+\tau _2\right) \omega _3-\tau _3, \qquad\qquad x_2 \mapsto 2 \left(\tau _1+\tau _2\right) \omega _1+4 \tau _0,
\end{multline*}
\begin{multline*}
 y_0 \mapsto Q \left(\tau _1+\tau _2\right) \tau _4+\tau _4 \left(-\tau _2 \omega _1^2+\tau _4 \omega _2+3 \tau _3
   \omega _3\right)-\tau _0 \left(2 \tau _4 \omega _1+\left(\tau _1+\tau _2\right) \omega _3+3 \tau
   _3\right)\\
 +\tau _1 \left(\tau _2 \left(\omega _2-2 \omega _1 \omega _3\right)-\tau _4 \omega _1^2\right),
\end{multline*}
\begin{multline*}
 y_1 \mapsto Q \left(\tau _1-\tau _2\right){}^2-4 \left(\tau _1+\tau _2\right) \tau _0 \omega _1-\left(\tau _1+\tau
   _2\right){}^2 \omega _1^2+4 \tau _3 \tau _4 \omega _1+2 \left(\tau _1+\tau _2\right) \tau _3 \omega _3-4 \tau
   _0^2-3 \tau _3^2+\tau _4^2+\tau _1 \tau _2,
\end{multline*}
\begin{multline}\label{special2bi}
 y_2 \mapsto \tau _1^2 \left(-\omega _2\right)-\tau _1 \left(\tau _4-2 \tau _2 \omega _2\right)+2 \left(\tau _1+\tau
   _2\right) \tau _3 \omega _1-\tau _2 \left(\tau _2 \omega _2+\tau _4\right)+4 \tau _0 \tau _3,\\
\end{multline}
where $Q$ is the Casimir \eqref{bi-casimir} of the Bannai-Ito algebra. Under the map $\psi$ defined in \eqref{psi-map} and \eqref{special2bi} with the parameters as in \eqref{special1bi}, the central element $\Lambda$ given in \eqref{hbi-casimir} is mapped to the Casimir of the Bannai-Ito algebra, up to a central element and a scaling, such that
\begin{align*}
 \psi:\Lambda \longmapsto u\, Q + v,
\end{align*}
where
\begin{align*}
 u = \tau _4^2+\tau _1 \tau _2, \qquad v = -2 \tau _0 \left(\tau _1 \omega _1+\tau _2 \omega _1-\tau _4 \omega _3\right)+\tau _1 \tau _4 \omega_2+\tau_4\left(\tau _2 \omega _2-\tau _4 \omega _1^2\right)-\tau _1 \tau _2 \left(\omega_1^2+\omega_3^2\right)-3\tau_0^2.
\end{align*}

\subsection{The Heun-Bannai-Ito operator}
The generalized Heun operator $W$ of the Bannai-Ito type can be introduced as the generic first order reflection operator in the infinite dihedral group $D_\infty$ that has the degree raising property:
\begin{align}\label{heun-property}
W p_{n}(x) \longmapsto q_{n+1}(x),
\end{align}
for $p_{n}$ and $q_{n}$ arbitrary polynomials of degree $n$ and $n+1$, respectively. Consider the general reflection operator $W$ specified by
\begin{align}\label{hbi}
 W = A_1(x)R_1 + A_2(x)R_2 + A_0(x) I.
\end{align}
Acting on the first three monomials in $x$ and demanding that \eqref{heun-property} holds fully determines the form of the coefficients $A_i(x)$ for $i=0,1,2$. One has
\begin{equation}\label{hbi-on-monomials}
 \begin{aligned}
 p_1(x) &\equiv W \cdot 1 = A_1(x) + A_2(x)+A_0(x),\\
 p_2(x) &\equiv W \cdot x = x (A_0(x)-A_1(x))-(x+1) A_2(x),\\
 p_3(x) &\equiv W \cdot x^2 = x^2 (A_0(x)+A_1(x))+(x+1)^2 A_2(x),
 \end{aligned}
\end{equation}
where $p_1(x)$, $p_2(x)$ and $p_3(x)$ are arbitrary polynomials of first, second and third degree, respectively. One solves easily \eqref{hbi-on-monomials} for the coefficients $A_i(x)$ to obtain
\begin{align}\label{hbi-coef}
\begin{aligned}
A_0(x) &= \frac{p_3(x)+(2x+1)p_2(x)+x(x+1)p_1(x)}{2x(2x+1)},\\
A_1(x) &= \frac{x(x+1)p_1(x)-p_2(x)-p_3(x)}{2x},\\
A_2(x) &= \frac{p_3(x)-x^2 p_1(x)}{2x+1}.
\end{aligned}
\end{align}
From the above, one sees that the coefficients of the polynomials $p_1(x)$, $p_2(x)$ and $p_3(x)$ constitute a parametrization of the Heun-Bannai-Ito operator. We will denote these nine parameters as follows
\begin{align}\label{hbi-param}
 p_1(x) = \sum\limits_{i=0}^1 p_1^{(i)} x^i, \qquad p_2(x) = \sum\limits_{i=0}^2 p_2^{(i)} x^i, \qquad p_3(x) = \sum\limits_{i=0}^3 p_3^{(i)} x^i.
\end{align}
For the Heun-Bannai-Ito operator to act on the finite Bannai-Ito grid $x_{s}$ given in \eqref{bi-grid}, additional constraints exist on the parameters. Depending on the truncation conditions, one has
\begin{align}
A_{1}(x) &\propto (x-\rho_{j}),\, A_{2}(x) \propto (x-r_{i}+1/2), & &\text{for }N\text{ even,} & 2(r_{i}+\rho_{j}) &= N+1, & i,j&=1,2,\nonumber\\
A_{1}(x) &\propto (x-\rho_{1})(x-\rho_{2}), & &\text{for }N\text{ odd,} & 2(\rho_{1}+\rho_{2}) &= -N-1,&&\nonumber\\
A_{2}(x) &\propto (x-r_{1}+1/2)(x-r_{2}+1/2), & &\text{for }N\text{ odd,} & 2(r_{1}+r_{2}) &= N+1.&&\nonumber\\
&&&&&&&\label{hbi-truncate}
\end{align}
These conditions can be expressed on the parameters \eqref{hbi-param} of the Heun-Bannai-Ito operator as the following constraints. For any constants, $a$ and $b$, one has that
\begin{align}
A_{1} &\propto (x-a) \implies p_{3}^{(0)} = a^{3}(p_{1}^{(1)}-p_{3}^{(3)})+a^{2}(p_{1}^{(0)}-p_{3}^{(2)})-a\, p_{3}^{(1)}\nonumber\\
A_{2} &\propto (x-b) \implies p_{2}^{(0)} = b^{3}(p_{1}^{(1)}-p_{3}^{(3)})+b^{2}(p_{1}^{(0)}+p_{1}^{(1)}-p_{2}^{(2)}-p_{3}^{(2)})+b (p_{1}^{(0)}-p_{2}^{(1)}-p_{3}^{(1)})-p_{3}^{(0)},\label{hbi-cons-param1}
\end{align}
and, conjunctly with the above, one also has
\begin{align}
A_{1} \propto (x-b) \implies p_{3}^{(1)} &= (a^{2}+b^{2})(p_{1}^{(1)}-p_{3}^{(3)})+a b (p_{1}^{(1)}-p_{3}^{(3)})+(a+b)(p_{1}^{(0)}-p_{3}^{(2)}),\nonumber\\
A_{2} \propto (x-a) \implies p_{2}^{(1)} &= (a^{2}+b^{2})(p_{1}^{(1)}-p_{3}^{(3)})+a b (p_{1}^{(1)}-p_{3}^{(3)})+(a+b)(p_{1}^{(0)}+p_{1}^{(1)}-p_{2}^{(2)}-p_{3}^{(2)})\nonumber\\
&\qquad\qquad\qquad\qquad\qquad\qquad\qquad\qquad\qquad\qquad\qquad\qquad +p_{1}^{(0)}-p_{3}^{(1)}.\label{hbi-cons-param2}
\end{align}
Using two of the above four constraints on the parameters, one can satisfy any case of the truncation conditions displayed in \eqref{hbi-truncate}. Thus, when constrained to the Bannai-Ito grid \eqref{bi-grid}, the Heun-Bannai-Ito operator has seven free parameters amongst those of \eqref{hbi-param}.

\subsection{Tridiagonalization in the Bannai-Ito algebra}
The Heun-Bannai-Ito operator can be obtained from the tridiagonalization procedure applied to the Bannai-Ito bispectral operators. Consider the following generic $W \in \mathcal{B}$
\begin{align}\label{bi-tridiag}
W =\tau_{1} B_1B_2 + \tau_{2} B_2B_1 + \tau_{3} B_1 + \tau_{4} B_2 + \tau_{0} I.
\end{align}
In the realization \eqref{bi-real}, it can be seen by direct calculations that the above operator takes the form of the Heun-Bannai-Ito operator \eqref{hbi}. Thus, the Heun-Bannai-Ito operator is one of the generators in the realization of the Heun-Bannai-Ito algebra constructed from the concatenation of the embedding map $\psi$ defined in \eqref{psi-map} with the realization given in \eqref{bi-real}.

Once the Bannai-Ito grid $x_{s}$ is specified as in \eqref{bi-grid}, the realization \eqref{bi-real} admits two free parameters. Thus, as the definition \eqref{bi-tridiag} for $W$ introduced five additional parameters, this realization of $W$ has seven free parameters, as was the case for the Heun-Bannai-Ito operator in \eqref{hbi-param} with \eqref{hbi-truncate}, \eqref{hbi-cons-param1} and \eqref{hbi-cons-param2}. In this case, the parameters of the Heun-Bannai-Ito operator as given by \eqref{hbi-param} can be given in terms of those of the realization \eqref{bi-real} of the Bannai-Ito algebra together with those of the tridiagonalization \eqref{bi-tridiag}. One obtains
\begin{align*}
 p_3{}^{(3)} &=\tau _1 \left(2 \rho _1+2 \rho _2-2 r_1-2 r_2+5\right)+\tau _2 \left(-2 \rho _1-2 \rho _2+2 r_1+2 r_2-7\right)+2 \tau_3,\\
 p_3{}^{(2)} &=\frac{1}{4} \big(2 \rho_1 \tau_2+16 \rho_1 \rho_2 \tau_2+2 \rho_2 \tau_2+4 \rho_1 \tau_4+4 \rho _2 \tau _4+\tau _1 \left(2 \rho _1+2 \rho _2-18 r_1-18 r_2+21\right)\\
 & \qquad\qquad\qquad +22 r_1 \tau_2-16 r_1 r_2 \tau _2+22 r_2 \tau _2-4 r_1 \tau _4-4 r_2 \tau _4+4 \tau _0-31 \tau _2+2 \tau _3+10 \tau _4\big),\\
p_3{}^{(1)} &=\left(-3 r_2+r_1 \left(4 r_2-3\right)+2\right) \tau _1+\left(r_1 \left(5-4 r_2\right)+5 r_2-4\right) \tau _2-2 \left(r_1+r_2-1\right) \tau _4,\\
p_3{}^{(0)} &=\frac{1}{4} \left(2 r_1-1\right) \left(2 r_2-1\right) \left(\tau _1-3 \tau _2+2 \tau _4\right),\\
p_2{}^{(2)} &=\tau _1 \left(-2 \rho _1-2 \rho _2+2 r_1+2 r_2-3\right)+\tau _2 \left(2 \rho _1+2 \rho _2-2 r_1-2 r_2+5\right)+2 \tau _3,\\
p_2{}^{(1)} &=\frac{1}{4} \big(-2 \rho _1 \tau _2-2 \rho _2 \tau _2-4 \rho _1 \tau _4-4 \rho _2 \tau _4+\tau _1 \left(-2 \rho _1+16 \rho _1 \rho _2-2 \rho _2+10 r_2-2 r_1 \left(8 r_2-5\right)-7\right)\\
& \qquad\qquad\qquad -14 r_1 \tau _2-14 r_2 \tau _2+4 r_1 \tau _4+4 r_2 \tau _4+4 \tau _0+13 \tau _2+2 \tau _3-6 \tau _4\big),\\
p_2{}^{(0)} &=\rho _1 \rho _2 \left(\tau _1+\tau _2+2 \tau _4\right)-\frac{1}{4} \left(2 r_1-1\right) \left(2 r_2-1\right) \left(\tau _1-3 \tau _2+2 \tau _4\right),\\
p_1{}^{(1)} &=\tau _1 \left(2 \rho _1+2 \rho _2-2 r_1-2 r_2+1\right)+\tau _2 \left(-2 \rho _1-2 \rho _2+2 r_1+2 r_2-3\right)+2 \tau_3,\\
p_1{}^{(0)} &=\frac{1}{4} \big(2 \rho _1 \tau _2+16 \rho _1 \rho _2 \tau _2+2 \rho _2 \tau _2+4 \rho _1 \tau _4+4 \rho _2 \tau _4+\tau _1 \left(2 \rho _1+2 \rho _2-2 r_1-2 r_2+1\right)+6 r_1 \tau _2\\
& \qquad\qquad\qquad -16 r_1 r_2 \tau _2+6 r_2 \tau _2-4 r_1 \tau _4-4 r_2 \tau _4+4 \tau _0-3 \tau _2+2 \tau _3+2 \tau _4\big).
\end{align*}

\newpage
\section*{Conclusion}

The recently introduced \cite{Grunbaum2018} notion of the algebraic Heun operator enables the construction of generalized Heun operators from a bispectral pair of operators. In particular, as each polynomial families in the Askey scheme is associated to a bispectral problem, a corresponding generalized Heun operator can be constructed. Furthermore, paralleling the algebraic approach to the Askey scheme, one is lead to the study of algebraic structures that encode the properties of these generalized Heun operators. This paper examined the Racah and Bannai-Ito cases.

The Heun-Racah operator was first constructed as the most general operator on the Racah grid satisfying the Heun property of sending polynomials to polynomials one degree higher. This operator could be identified with the algebraic Heun operator of the Racah type in the canonical realization of the Racah algebra. The Heun-Racah algebra associated to the Racah polynomials was subsequently introduced. This algebra was defined in a generic presentation and a central element was identified. The association with the Racah algebra was made explicit by the identification of a map that embeds a specialization of the Heun-Racah algebra, obtained from conditions on the parameters, as a subalgebra of the Racah algebra. This embedding effectively maps the central element to the Casimir operator of the Racah algebra. Moreover, using this embedding, a realization of the specialized Heun-Racah algebra is induced by the canonical realization of the Racah algebra.

As the Racah polynomials are at the top of the Askey scheme, the algebraic structure that result from the construction of the algebraic Heun operator does not correspond to an algebra associated to polynomials of the Askey scheme. This motivates further examination of the Heun-Racah algebra as a new algebraic structure. Moreover, in view of the limit $q \to 1$ that relates the Askey-Wilson algebra to the Racah algebra, one would expect a similar limit that relates the results in \cite{Baseilhac2018} with the results presented here.

An analogous examination was made for the Bannai-Ito case. The Heun-Bannai-Ito algebra was first introduced abstractly and a specialization with conditions on the parameters was shown to embed in the Bannai-Ito algebra. In the canonical realization of the Bannai-Ito algebra, this specialization was demonstrated to be realized in terms of the associated Heun-Bannai-Ito operator.

In view of the embedding of the Racah algebra in the Bannai-Ito algebra presented in section \ref{bannai-ito}, one could ask if the relation between these two algebraic structures is reflected in the associated Heun algebras. Indeed, this paper has illustrated the following maps between these algebraic structures
\begin{center}
\begin{tikzcd}
&\mathcal{R} \ar[rd,"\chi"]&\\
\mathcal{HR} \ar[ru,"\phi"] \ar[rr,"\chi\circ\phi",dashed] \ar[rrdd,"\Upsilon",dashed] &  & \tilde{\mathcal{R}} \ar[dd,"\theta"]\\
&&\\
\mathcal{HB} \ar[rr,"\psi"] &  & \mathcal{B},
\end{tikzcd}
\end{center}
where $\chi,\, \phi,\, \theta$ and $\psi$ are defined in \eqref{to-eq-racah}, \eqref{phi-emb}, \eqref{r-to-bi} and \eqref{psi-map}, respectively and $\Upsilon = \theta\circ\chi\circ\phi$. Furthermore, it can be shown that
\begin{multline*}
\Upsilon (W_{\mathcal{HR}}) = b_{1} \psi(W_{\mathcal{HB}})^{2} + b_{2} \{ \Gamma, \psi(W_{\mathcal{HB}}) \}+[\psi(X_{\mathcal{HB}}),\psi(W_{\mathcal{HB}})] + b_{4} [ \Gamma,\psi(W_{\mathcal{HB}}) ] + b_{5} Q\\
b_{6} \{ \psi(X_{\mathcal{HB}}), \psi(W_{\mathcal{HB}}) \} + b_{7} \{ \psi(X_{\mathcal{HB}}) , \Gamma \} + b_{8} \psi(W_{\mathcal{HB}}),
\end{multline*}
where the subscripts $\mathcal{HR}$ and $\mathcal{HB}$ denote, respectively, generators in the Heun-Racah and Heun-Bannai-Ito algebra and where $b_{i}$ for $1\leq i\leq 8$ are coefficients. Finding a specialization of the Heun-Racah algebra that embeds in the Heun-Bannai-Ito algebra would enhance the association between these new algebraic structures and the algebraic structures of the Askey scheme.

\newpage
\section*{Acknowledgements}
The research of G.B. is supported in part by a scholarship of the Natural Sciences and Engineering Research Council of Canada (NSERC). N.C. warmly thanks the Centre de Recherches Mathématiques (CRM) for hospitality and support during his visit to Montreal in the course of this investigation. The research of L.V. is supported in part by a Discovery Grant from NSERC. The work of A.Z. is supported by the National Science Foundation of China (Grant No. 11771015).

\section*{Data availability}
The data that supports the findings of this study are available within the article.

\bibliographystyle{abbrv}
\bibliography{heun.bib}

\end{document}